\pdfoutput=1
\documentclass[fleqn,usenatbib]{mnras}
\usepackage{newtxtext,newtxmath}

\usepackage[T1]{fontenc}
\usepackage{ae,aecompl}
\usepackage[title]{appendix}

\usepackage{graphicx}	
\usepackage{amsmath}	
\usepackage{amssymb}	
\usepackage{epstopdf}
\usepackage{epsfig}

\newcommand{\fractal}{\%\mathrm{Frac}}

\newcommand{\dss}{$\delta$ Scuti stars }

\newcommand{\eqn} [1] {
\begin{equation}
#1
\end{equation}}

\newcommand{\eqna} [1] {
\begin{eqnarray}
#1
\end{eqnarray}}

\title[A fractal analysis application of the pre-whitening technique to \dss  time series]{ A fractal analysis application of the pre-whitening technique to \dss  time series}
\author[S. de Franciscis et al.]{
S. de Franciscis,$^{1}$\thanks{E-mail: sebas@iaa.es},
J. Pascual-Granado$^{1}$,
J.C. Su\'arez$^{2,1}$,
A. Garc\'ia Hern\'andez$^{2}$,
\newauthor{R. Garrido$^{1}$,
M. Lares-Martiz$^{1}$,
J.R. Rod\'on$^{1}$.}
\\
$^{1}$Instituto de Astrof\'isica de Andaluc\'ia (IAA-CSIC). Glorieta de Astronom\'ia s/n, 18008, Granada. Spain\\
$^{2}$ Dept. F\'isica Te\'orica y del Cosmos. Univ. Granada (UGR). Av. de Fuente Nueva S/N. 18071, Granada. Spain}

\date{Accepted XXX. Received YYY; in original form ZZZ}

\pubyear{2018}

\hypersetup{draft}
\begin{document}
\label{firstpage}
\pagerange{\pageref{firstpage}--\pageref{lastpage}}
\maketitle

\begin{abstract}

Fractal fingerprints have been found recently in the light curves of several \dss\ observed by CoRoT satellite. This sole fact might pose a problem for the detection of pulsation frequencies using classical pre-whitening techniques, but it is also a potentially rich source for information about physical mechanisms associated to stellar variability.

Assuming that a light curve is composed by a superposition of oscillation modes with a fractal background noise, in this work we applied the Coarse Graining Spectral Analysis (CGSA), an FFT based algorithm, which can discriminate in a time series the stochastic fractal power spectra from the harmonic one. 

We have found that the fractal background component is determining the frequency content extracted using classical pre-whitening techniques in the light curves of \dss. This might be crucial to understand the amount of frequencies excited in these kind of pulsating stars. Additionally, CGSA resulted to be relevant in order to extract the oscillation modes, this points to a new criterion to stop the pre-whitening cascade based on the percentage of fractal component in the residuals.

\end{abstract}



\section{Introduction}\label{sec:intro}
 So far, fractal analysis has been applied with success to characterize light curves of solar-like stars \citep{NewSunsI,NewSunsIII} as well as light variations of M dwarfs \citep{NewSunsIV}.\\  
More recently we introduced different tools from the field of fractal analysis to light curves of  \dss\, with the aim to detect any fractal fingerprint  that may help to better understand their pulsational content \citep{de2018fractal}. Indeed, \dss\ might exhibit rather complex oscillation spectra due to the presence of a thin outer convective zone together with a rapid rotation. These properties have been also used to test theories of angular momentum transport and chemicals \citep{Goupil05}. \par
The \dss\ are intermediate-mass ($1.5 - 3\,\mathrm{M}_\odot$), main-sequence A-F type pulsators, whose oscillations are maintained from the varying ionization of helium ($\kappa$ mechanism), sharing physical properties with $\gamma$ Doradus stars which seem to have g modes excited by the convective flux blocking as described for the first time by \citet{Guzik2000}.\\
In addition, ultra-precise data from CoRoT  (Convection Rotation and planetary Transits, \citealt{Baglin06}) and Kepler \citep{Gilli10} have revealed that after a pre-whitening process the remaining power spectra is higher than the expected from the high frequency region, where the noise is supposed to be the dominant component. The region where the power is higher, forms a  plateau \citep[see e.g. figure 1 in][]{Poretti09HD5084}.\\
It has been suggested a possible granulation background signal to be the origin of these light variation \citep{KallingerMat10,Balona11}. A thin outer convective layer  was claimed to be the cause of the excitation of solar-like oscillations in \dss \citep{Antoci11}.\\
Alternative explanation have been explored by \citet{BarceloForteza15}, or the existence of a magnetic field by \citet{NeinerLampens15}. Also, a possible explanation  can be the large number of chaotic modes \citep{LignieresGeor09} predicted in non-spherical stellar model  due to high rotation \citep[recently confirmed by][]{BarceloForteza17}.\\
In a recent work \citep{Javi18Prew} a dataset of CoRoT \dss light curves has been analyzed using SIGSPEC, a pre-whitening algorithm that, using a rigorous statistical treatment of maxima in a Discrete Fourier Transform (DFT),  identifies the harmonic components of a given light curve corresponding to a pulsating star.\\ 
The paper is organized as follows: in Section~\ref{sec:safine} we specify the mathematical framework in which the fractal analysis is performed; then, Section~\ref{sec:corotdata} describes the observational data selected, while Section~\ref{sec:methods} outlines the methodology followed; discussion of the results is presented in Section~\ref{sec:results}, and conclusions are summarized in Section~\ref{sec:conclusions}.

\section{Self-affine time series}\label{sec:safine}
Fractal spatial and temporal dynamics emerges in the context of critical phenomena, i.e. at  the frontier between the ordered and disordered phases of a system, where the strength of stochastic fluctuations and deterministic dynamic are in a sort of dynamical equilibrium \citep{Chandler87,Dickman00}.
Physical systems in a critical state are characterized by scale invariance, i.e. they have the same appearance at any spatial and temporal scale, and the characteristic functions of the system, as correlation functions or probability distributions, have a typical power law shape, i.e. the so called scale free distribution: $f(x)=x^{\zeta}$ \citep{Stanley72}.\\
In stellar physics fractal fingerprints have been found in statistical, power law distributed, observables as in perimeter/area correlations \citep{SolGran86} and size and lifetime distributions of solar granules \citep{Lemme17}, in sunspot number and area variability \citep{zhou14, drozdz}, in magnetospheric substorms, auroras and flares \citep{SOCAstro12}, and finally in light curves from pulsating stars \citep{JaviFrac11, NewSunsI, NewSunsIII, NewSunsIV}.\\
In asteroseismology we are not working with spatially extended objects, and we can only extract information on stellar structures and dynamic from their integrated light curves. Thus we need to work with a generalization of fractal from geometrical objects to time series, which lead to the property of Self-Affinity:  time series $y(t)$ is self-affine if it has the following inhomogeneous scaling relation \citep{Turco}:
\eqn{y(t)=\lambda^{H}y(\lambda t) \label{uno}}
where H is the so-called Hausdorff exponent, characterizing long term correlations and the type of self-affinity in time series. Equation \ref{uno} has to be taken in statistical meaning, so that the scaling relationship holds when one performs appropriate measures on mean values over pairs of points at the same distance or over equal length subseries or windows etc.\\
In this work we made the assumption  that the light curve of \dss can be considered as the sum of  a self-affine background signal $y(t)$ and a harmonic function, representing the pulsating modes, i.e. :
\eqn{f(t)=y(t)+\sum_{i=1}^{n}A_{i}\sin(2\pi\omega_it+\phi_i);\label{dos}}
here $A_{i}$, $\omega_{i}$ and $\phi_{i}$ are respectively amplitude, frequency and phase of the $i-th$ modes. In the following section we will show some results on the study of a sample of the light curves of $\delta$ Scuti stars in searching for any fractal signature.

\begin{table*}
 	\caption{Coarse Grained Spectral Analysis $\%Frac$ for the sample of 15 CoRoT \dss\ studied. Calculated on the original time series gap-filled with MIARMA algorithm}

    \label{tab:CGSA_All}
	\begin{tabular}[!h]{lcccccccccc}
ID & CGSA $\%Frac$  &  SpT & Teff (K) & $\log g$ & $M_{v}$ & $\mathrm{v}\cdot\sin i$ ($\mathrm{km}\cdot\mathrm{s}^{-1}$) & PulsT & Obs. time (d)\\
\hline
GSC00144-03031 & 0.031 & A8 & 7822$\pm$400 & 3.41$\pm$0.32 & & & HADS & 79.133\\ 
HD170699 & 0.151 & A2 & 7400 & 3.5 & 1.49 & 270 & $\delta$~Sct & 89.282\\ 
HD172189 & 0.336  & A6V & 7600$\pm$150 & 3.48$\pm$0.08 & 1.04 & 78$\pm$3 & $\delta$~Sct (EB) & 149.013\\ 
HD174532 & 0.070  & A2 & 6783$\pm$228 & 3.684 & 1.38 & 32 & $\delta$~Sct  & 26.239\\ 
HD174589 & 0.032 &  F3III & 7359$\pm$251 & 3.88$\pm$0.14 & 1.45 & 101 & $\delta$~Sct  & 26.168\\ 
HD174936 & 0.117 &  A2 & 8000$\pm$200 & 4.08$\pm$0.2 & 1.88 & 169.7 & $\delta$~Sct & 27.194\\ 
HD174966 & 0.008 &  A3 & 7555$\pm$50 & 4.21$\pm$0.05 & 1.95 &126.1$\pm$1.2 & $\delta$~Sct & 27.197 \\ 
HD181555 & 0.062 &  A5V & 7000$\pm$200 & 4.3$\pm$0.2 & -0.72 &200 &  $\delta$~Sct & 156.645\\ 
HD41641 & 0.067  &  A5V & 7561 & 3.71 & 1.92&28 &  $\delta$~Sct & 94.432\\ 
HD48784 & 0.224 &  F0 & 6990$\pm$140 & & 1.87&108 & Hybrid & 25.305 \\ 
HD49434 & 0.235 &  F1IV & 7632$\pm$126 & 4.43$\pm$0.2 & 2.74 &85.7$\pm$4.3 & Hybrid & 136.890\\ 
HD50844 & 0.055 &  A2 & 7400$\pm$200 & 3.6$\pm$0.2 & 1.31&58$\pm$2 & $\delta$~Sct  & 57.713\\ 
HD50870 & 0.019 &  A8III & 7660$\pm$250 & 3.68$\pm$0.25 & 1.67& 37.5$\pm$2.5  & $\delta$~Sct  & 114.413\\ 
HD51359 & 0.087  &  A5 & 6787$\pm$100 & & 0.89 &  & $\delta$~Sct  & 117.41\\ 
HD51722 & 0.033 &  A5 & 7051$\pm$100 & 3.544 & 1.13 & 127 & $\delta$~Sct  &117.375
\end{tabular}
\end{table*}

\section{CoRoT dataset and Frequency Extraction}\label{sec:corotdata}
In order to avoid spurious (instrumental) effects that may mimic the fractal behaviour in the stellar light curves we have selected a sample of stars observed by the CoRoT satellite, covering different observation times, and physical properties. This sample is the same as in \citet{Javi18Prew} and it is obtained by searching for $\delta$ Scuti stars in the Seismofield of CoRoT.\\
Among the sample of 15 stars, two have been identified as hybrid $\gamma$ Dor - $\delta$ Scuti stars, i.e. HD49434 \citep{ChapellierHD49434} and HD48784 \citep{BarceloForteza17}, while HD172189 is an eclipsing binary (EB) system \citep{Susana05}, and GSC0144-03031 is a HADS \citep{Poretti05},
and the rest are small amplitude oscillating stars.\\
Details of the physical parameters of the stars are shown in Table~\ref{tab:CGSA_All}). Their light curves have a total of $0.7-3.7\cdot10^5$ data points sampled at every 32 s. Details on the observing runs and other  observational characteristics can be found in \citet{Javi18Prew}.\\
One possible source of misleading fractality might be the presence of gaps in the data, which unfortunately occur at every orbit around the earth, due to the satellite passage through the South Atlantic Anomaly. To overcome this problem the gaps in the light curves were filled by an MIARMA algorithm, which preserve the original frequency content of the signal \citep{Javi15miarma}.\\
Then once obtained evenly-spaced time series for every target, they were analyzed using the widespread numerical recipe SIGSPEC \citep{Reegen07}. This algorithm estimates the frequency content of the light curves by giving a sequence of frequencies for which their detection is statistically significant. The aforementioned process is very similar to the CLEAN algorithm used in radioastronomy, which performs a subtraction of frequencies in the time domain identified as pre-whitening.\\
The frequency detection is performed through a significance spectrum based on the false alarm probability of each peak appearing in the power spectrum. The iterative sequence stops when a significance threshold is reached (by default $sig =5.0$, that is $\approx S/N=4$). Here, we follow the procedure as in \citet{Javi18Prew} using the same set of free parameters. Among others, we choose a frequency interval of $2-100$ $d^{-1}$.\\
Special considerations arose around two targets: the eclipsing binary HD172189 and the HADS GSC00144-03031.
The transits of the eclipsing binary star HD172189 introduce nonlinear effects that could bias the frequency analysis performed by classical pre-whitening techniques. Additionally, during the preparatory phase of this work we obtained results suggesting that the standard frequency analysis procedure for the HADS GSC00144-03031 needs to be revised to include proper frequency combinations and harmonics in the pre-whitening (Lares-Martiz et al. 2018, in prep.).\\
However, we decided to include both HD172189 and GSC00144-03031 in our analysis to verify if their results differ significantly from the other stars.

\section{Fractal analysis techniques}\label{sec:methods}
Fractal nature of  \dss light curves have been previously studied and detected in some stars of our sample in \citep{de2018fractal},  in which  emerges a clear power law dependence in Rescaled Range analysis (RR) and Fourier power spectra.\\
Here we perform a CGSA analysis on the residual time series output of our light curves sample. The Coarse Graining Spectral Analysis splits in a time series the self-affine component and the harmonic one, giving as output the percentage of (stochastic) fractal power in time series \citep{Yama}. CGSA is based on the consideration that in a self-similar time series the FFT phases follows a uniform distribution $\Theta_k\in \left[0, 2\pi \right]$. We consider the original time series $y(i)$ %
and the series obtained by scaling $y(i)$ by a factor $2$ and $1/2$:
\eqna{
	y_2 &=& \{ y(2),y(4),y(6), \ldots\} \\
	y_{1/2}&=&\{y(1),y(1),y(2),y(2),\ldots \} . 
}
Next we cut those series in  $N_s$ partially overlapping subsets, each one having size the $90\%$ of the total length \footnote{$N_s$ is an additional free parameter.}. 
For each window $m$ we compute the auto-power spectrum $S_{yy,m}$ and the cross-power spectrum\footnote{Cross-power spectrum is defined as the Fourier transform of the cross-correlation function, i.e. $S_{XY}(k)=\sum_{n}\sum_{n'}X(n)Y(n+n')e^{ikn}$.} between the original series and the rescaled ones, i.e. $S_{yy_2,m}$ and $S_{yy_{\frac{1}{2}},m}$. 
If $y(i)$ is constituted by a sum of a few harmonics with fixed phase relationship it is possible to exploit the phase difference between windows $m-2$ and $m-1$ to orthogonalize $S_{yy_z,m}$ where $z \in \{1/2, 2\}$, with the rotating factor 
\eqn{S_{yy_z,m}^{o}(k)=S_{yy_z,m}(k)e^{-i\left[\pi/2-(\Theta_{m-1,k}-\Theta_{m-2,k})\right]}}
The residuals of such orthogonalization process are non zero in self-affine series, because any rescaled harmonic will find its counterpart in the original series and the phase relationships are always randomly distributed. Taking advantage of Schwartz's inequality we can calculate the fractal module cross correlations  
\eqna{
\langle || S_{yy_z,m-1}^{frac}(k) || \rangle_m &\equiv& \frac{\langle|| S_{yy_z,m-1}(k)\cdot S_{yy_z,m}^{o}(k) ||\rangle_m}{\langle S_{yy_z,m-1}(k)\rangle_m} \\
&\leq& \langle\mid\mid S_{yy_z,m}^{o}(k) \mid\mid \rangle_m.
}
Finally considering the possible distortions that could emerge by the finite size of the original series and the coarse graining of $y_{2}$ and $y_{1/2}$, we define the fractal power and the percentage of fractal power as
\eqna{
|| S^{frac}(k) || &\equiv& \sqrt{ || S_{yy_2}^{frac}(k) || \cdot || S_{yy_\frac{1}{2}}^{frac}(k) || }\\
\fractal &\equiv& \frac{\sum_{k}|| S^{frac}(k) ||}{\sum_{k}|| S(k) ||.}
}
SIGSPEC algorithm works essentially by extracting at each step the mode with the highest significance from the residual time series. In this way the sequence of modes are ordered by their power spectra amplitude. For each step of the pre-whitening, the CGSA is computed for the corresponding residuals, measuring the fractality percentage of each time series and measuring how the self-affine background weight evolves along the pre-whitening cascade.
\section{Results}\label{sec:results}
As can be seen in Figs. 1-3, where we study in details the first 150 pre-whitening steps, in the vast majority of cases already in the first tens steps the CGSA of the residuals increases monotonically, then they seems to reach an asymptotic value. Nevertheless extending the range of steps up to the whole pre-whitening output, up to hundreds and thousands of frequencies, the picture is very different, in the most of the cases the curves still grow, or grow and decrease with a long tail.\\
Figs. 4-6 shows the CGSA of the residual time series at every 50 steps of the pre-whitening in the order given by SIGSPEC. It is possible to observe that CGSA $\%Frac$ of the residual time series has different evolutions with the number of step $n$ depending from the star under examination. The increasing monotonic behaviour and the reaching of a plateau in $\%Frac(n)$ (figures 4 and 6) are a good indication of the mechanism by which SIGSPEC pre-whitening extracts step by step the modes, increasing the "fractality" of the resulting residual series, until the plateau, where all the oscillating modes have been already extracted. For those stars where $\%Frac(n)$ has a decreasing cue, showing a steep initial increase with a maximum of CGSA in the first 150-200 first steps followed by a decreasing tail, SIGSPEC appears to be extracting modes in residuals containing only background noise i.e. introducing a new artificial signal (figure 5). This explains why the pre-whitening process causes a cascade of spurious frequencies \citep{Balona2014}.\par
This shows that this new pre-whitening criteria is more robust for extracting the sinusoidal components of the light curves.\\
A first attempt to introduce a novel STOP criteria for SIGSPEC could be to study $\%Frac(n)$ and establish where it reaches the plateau. Nevertheless maximum value of $\%Frac$ depends on the self-affine exponent of noise signal \cite[see appendix A3 in:][]{de2018fractal}, and moreover local fluctuations are perturbing the study of the first derivative of those curves.\par
In table~\ref{tab:cut} we collect the values for the critical step in the pre-whitening cascade determined according to a CGSA based criterion. In order to reduce the fluctuation we applied to the CGSA curves a 5-point moving average filter. Then, several approaches were tried to find the best estimation of the critical step (i.e. stop criterion). It was found that most of the CGSA curves can be fitted optimally with a hyperbolic tangent function. Since this function has no critical points and their derivatives are smooth functions we finally used an epsilon around the limit value of CGSA to determine whether the critical step is reached. We have found empirically in our tests that $CGSA(N)-CGSA(i) < \epsilon = 0.02$ (where $CGSA(N)$ is the last value) is a robust value.\\
This procedure is valid only for those cases where CGSA is a strictly monotonically increasing function. If this function has a maximum and then decreases this points to an anomaly which is caused by the lack of convergence of the pre-whitening cascade to a uncorrelated Gaussian noise. We previously mentioned that an anomalous frequency analysis might be related to non-linearities introduced by a binary component of the stellar system (e.g. HD172189), by the non-linear response to the mechanism exciting pulsations in the star (e.g. GSC00144-03031) or by the spectrum modulation from the significant deformation of a fast rotating star (e.g. HD181555). This check proves the robustness of the method.\\
Finally regarding the last anomalous case HD51359, our study could be a hint showing that this star belongs to one of the aforementioned cases, but there is no clue in the literature about this, so complementary observations are necessary to confirm this point.\\
The number of frequencies extracted at the critical step shown in table~\ref{tab:cut} can be compared with the total number of frequencies extracted at the end of the pre-whitening cascade (Nf), i.e. when SIGSPEC stops, and the number of frequencies found in the literature. Notice that in most of the cases the number of frequencies extracted at the critical step is more similar to the value shown in the literature. However, the comparison with the literature must be done carefully since different criteria are used in different manuscripts, e.g. in \citet{Garcia13HD174966} 185 frequencies were extracted up to sig=10.0, but we have used the default value sig=5.0, which is more similar to the typical signal-to-noise ratio equal to 4. Therefore, the comparison with Nf is more reliable since we used a homogeneous procedure for the whole sample of stars. Our stop criterion gives a reduced set of frequencies containing from 16\% (HD 50870) to 89\% (HD 174532) of the frequencies extracted with a classical pre-whitening.\\
In order to better understand the CGSA behaviour with respect the pre-whitening step we build a controlled theoretical experiment, i.e. a toy-model of a typical \dss light curve. This model is based on our main assumption of eq. \ref{dos}, thus the light curve of HD174523 is built by summing sinusoidal functions with amplitudes, frequencies and phases obtained from the last pre-whitening step, i.e. the oscillation spectrum, together with a Gaussian white noise, with statistical properties (mean and standard deviation) similar to those given by light curve residuals.
The toy model representative of HD174523 has an analogue CGSA curves (Fig. \ref{CGSA_vs_step_Toy}), although  not identical. This is somewhat expected since the only source of fractal behaviour included in the model is the Gaussian white noise, whereas the observed light curves may have more (e.g. turbulence, convection, etc.). Nevertheless, A-D test performed on the frequency distributions of original and toy model shows that the distributions of toy model and data are compatible along the whole pre-whitening iterations \citep{Javi18Prew}. This evidence is consistent with previous results obtained for this star, since the critical step is more compatible with SIGSPEC than in other cases. An ongoing work is aimed to extend this study to the whole dataset.\\
Finally, several hypotheses are compatible with a fractal background signal: granulation, magnetic cycles, rotational effects, etc. In order to understand the physical origin of this component simulations from more detailed theoretical models should be performed. In summary, the self-affine component found in the light curves of the sample of stars studied here points to a revision of the stellar models applied to \dss.\\
The stop criterion suggested here for the pre-whitening procedure might be implemented easily in any frequency analysis package. In particular, CGSA will be implemented on a new version of SIGSPEC that avoids failing to stop when no harmonic component is left in the residuals.

\begin{table*}
	\caption{Critical values for the prewhitening process for each star analyzed. If the CGSA is increasing monotonically a critical step is reached when the CGSA fraction approximates an asymptotic value within a certain tolerance interval. Third column shows the total number of frequencies extracted at the end of the prewhitening cascade and the 4th column shows the number of frequencies extracted in the cited references.  (1) \citet{Hernandez09HD174936}, (2) \citet{Garcia13HD174966}, (3) \citet{BarceloForteza17}, (4) \citet{ChapellierHD49434}, (5) \citet{Poretti09HD5084}, (6) \citet{Mantegazza12HD50870}.}
	\label{tab:cut}
	\begin{tabular}[!h]{lcccc}
	Star & Step & CGSA & Nf & Literature\\
    \hline
    HD170699 & 2000 & 1.1677 & 3442 & - \\
    HD174532 & 850  & 1.0625 & 951 & - \\
	HD174589 & 100  & 1.0656 & 503 & - \\
    HD174936 & 400  & 1.0735 & 870 & $422^1$\\
    HD174966 & 400  & 1.0902 & 647 & $185^2$\\
	HD41641  & 2050 & 0.7900 & 2550 & - \\
    HD48784  & 250  & 1.0947 & 464 & $163^3$\\
    HD49434  & 850  & 0.7901 & 1612 & $840^4$\\
    HD50844  & 1450 & 1.0104 & 1746 & $1000^5$\\
	HD50870  & 400  & 0.8374 & 2484 & $734^6$\\  
	HD51722  & 300  & 0.7929 & 1693 & - \\
	\end{tabular}
\end{table*}

\begin{figure}
	\centering
		\includegraphics[width=9cm]{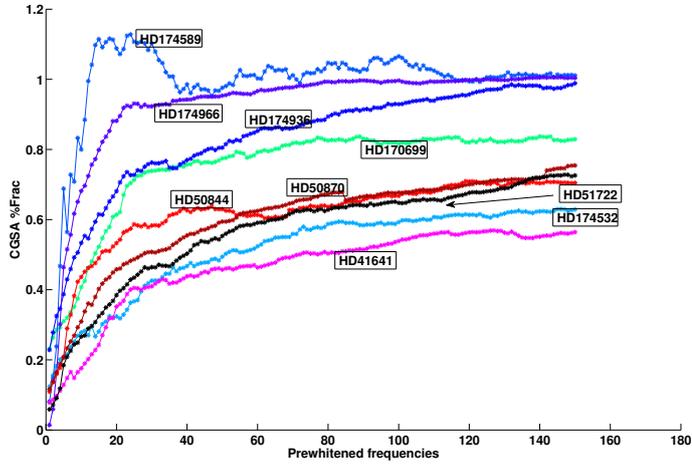}
	\caption{$\% Frac$ CGSA vs pre-whitening step for $\delta$-Scuti stars database under examination. First 150 steps are represented here. Color scheme is the same as in Figure 1.}
	\label{CGSA_vs_step_blue150}
\end{figure}
\begin{figure}
	\centering
		\includegraphics[width=9cm]{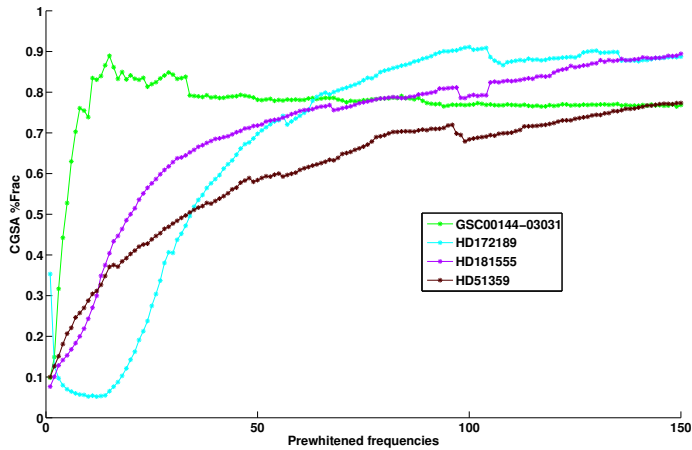}
	\caption{$\% Frac$ CGSA vs pre-whitening step for $\delta$-Scuti stars database samples with anomalous pre-whitening cascade. First 150 steps are represented here. Color scheme is the same as in Figure 2.}
	\label{CGSA_vs_step_red150}
\end{figure}
\begin{figure}
	\centering
		\includegraphics[width=9cm]{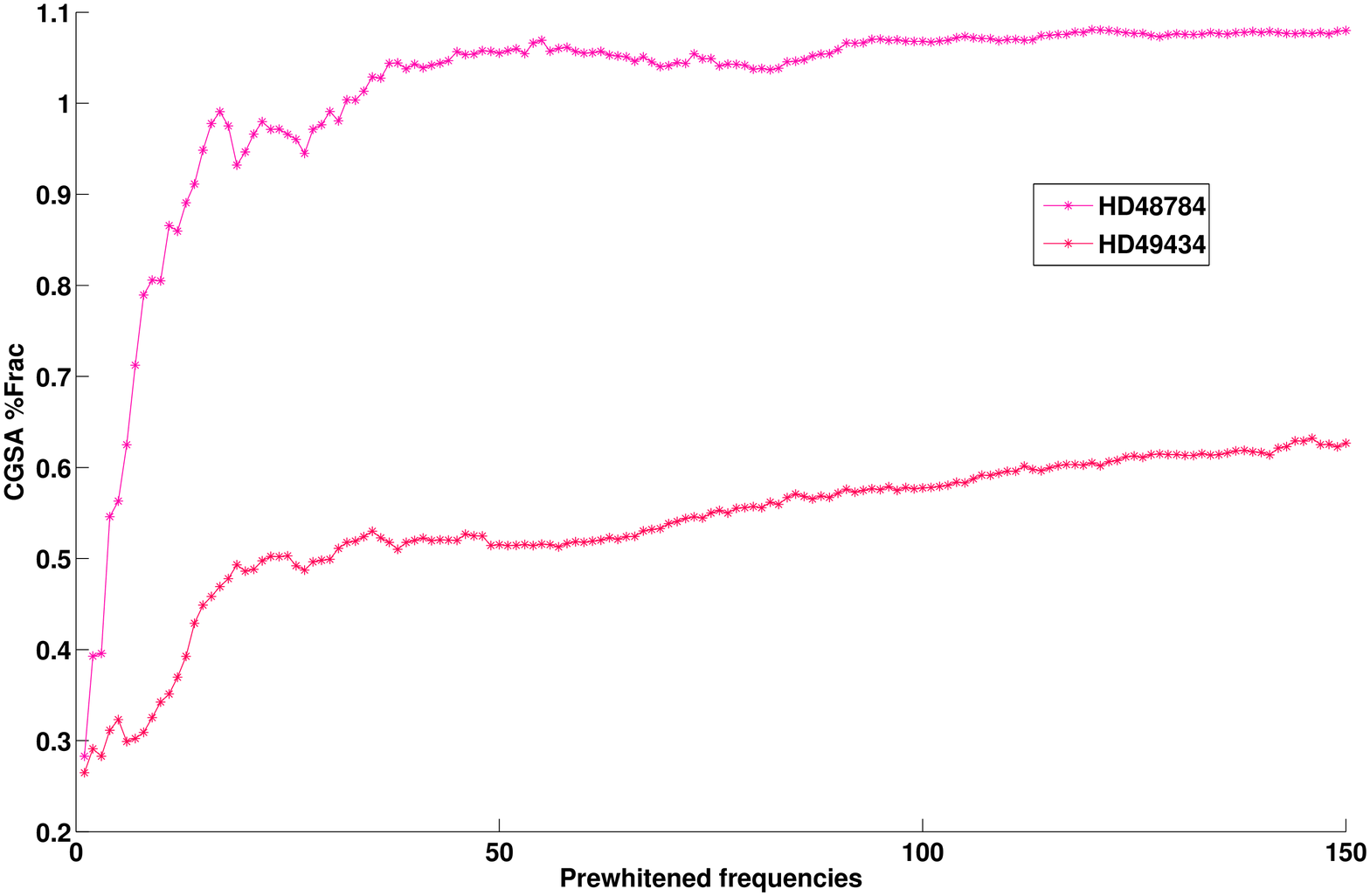}
	\caption{$\% Frac$ CGSA vs pre-whitening step for Gamma Doradus stars database samples. First 150 steps are represented here. Color scheme is the same as in Figure 3.}
	\label{CGSA_vs_step_green150}
\end{figure}

\begin{figure}
	\centering
		\includegraphics[width=9cm]{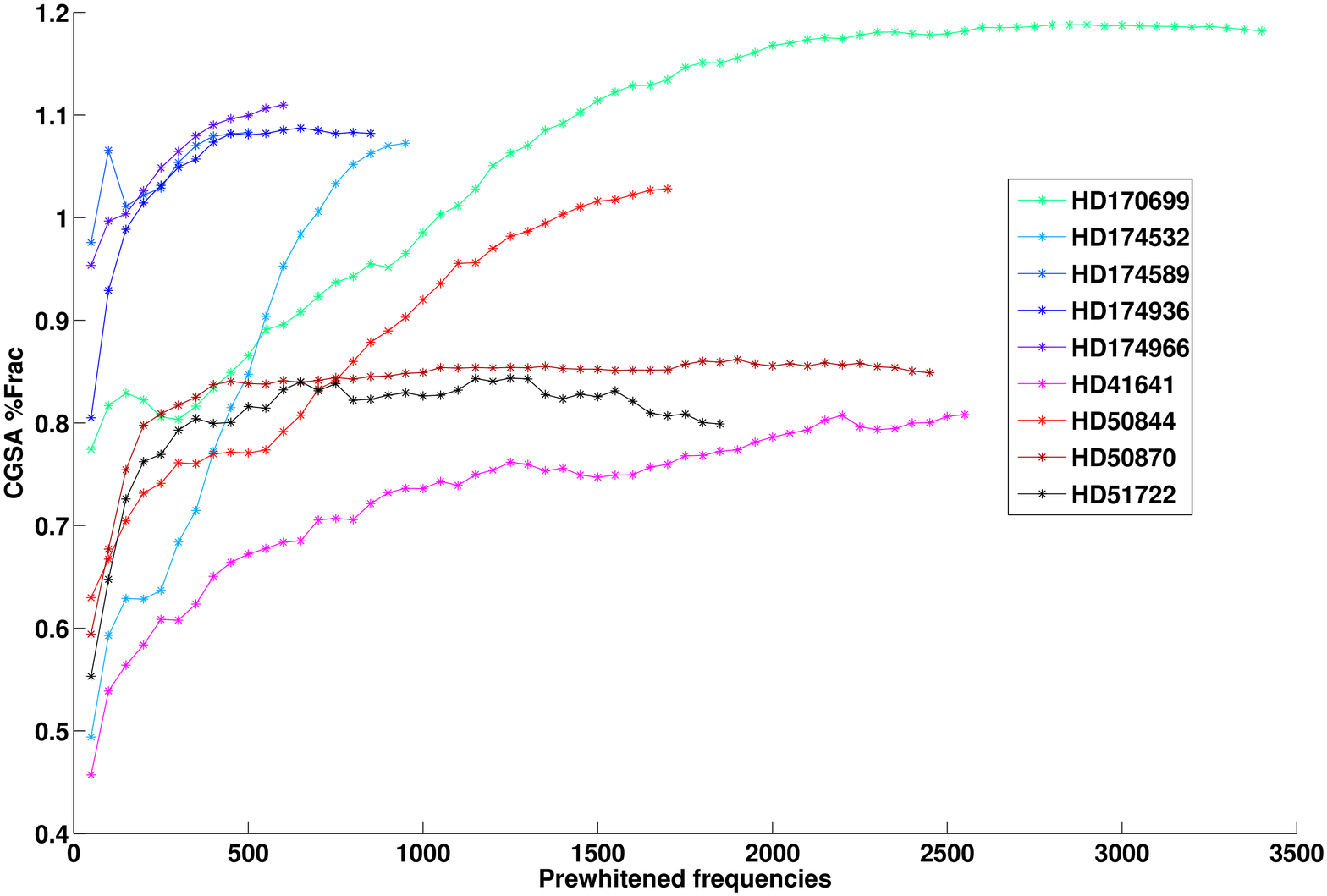}
	\caption{$\% Frac$ CGSA vs pre-whitening step for the $\delta$-Scuti stars samples. Points are separated by 50 pre-whitening steps.}
	\label{CGSA_vs_step_blue_All}
\end{figure}

\begin{figure}
	\centering
		\includegraphics[width=9cm]{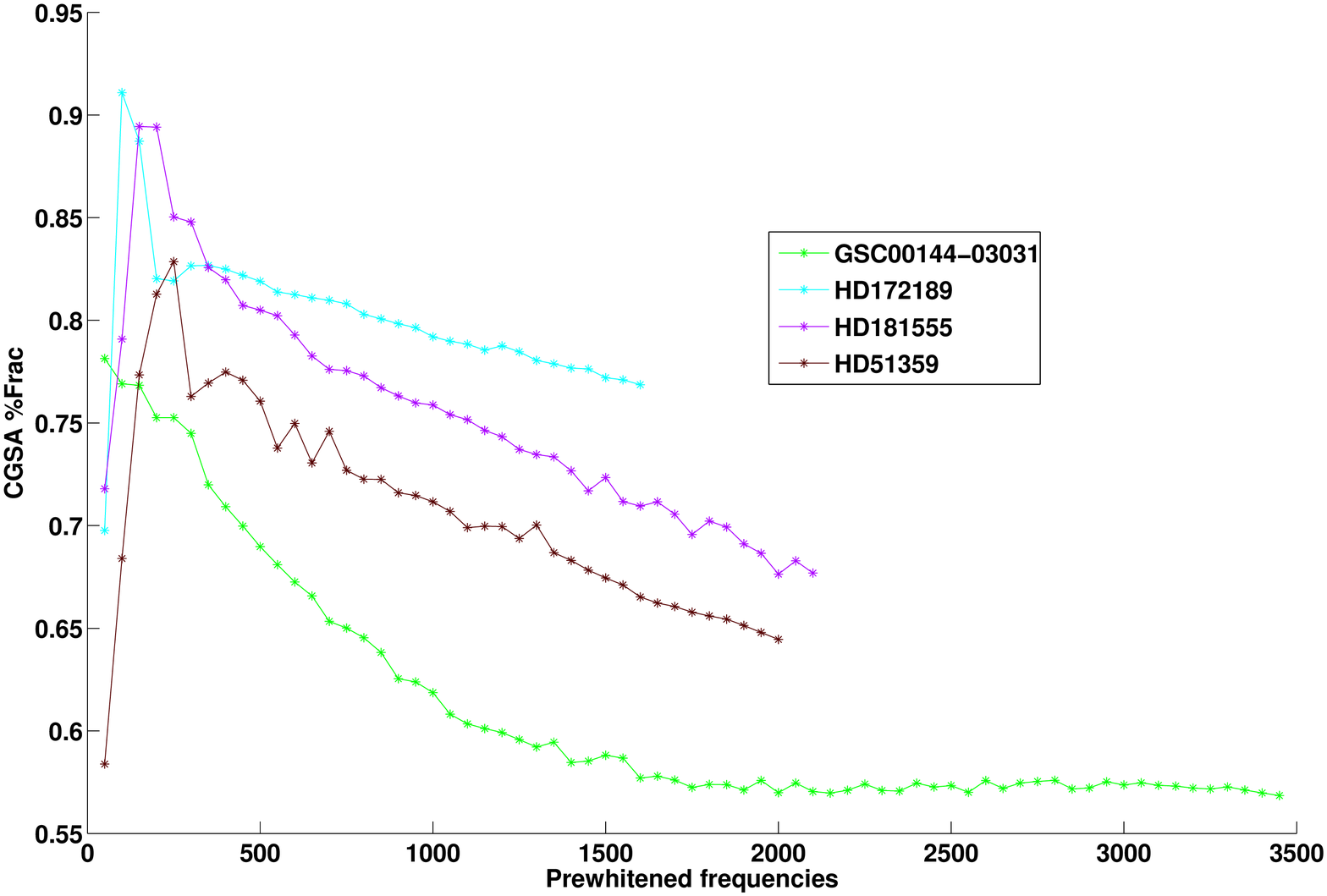}
	\caption{$\% Frac$ CGSA vs pre-whitening step for the $\delta$-Scuti stars samples with anomalous pre-whitening cascade. Points are separated by 50 pre-whitening steps.}
	\label{CGSA_vs_step_red_All}
\end{figure}

\begin{figure}
	\centering
		\includegraphics[width=9cm]{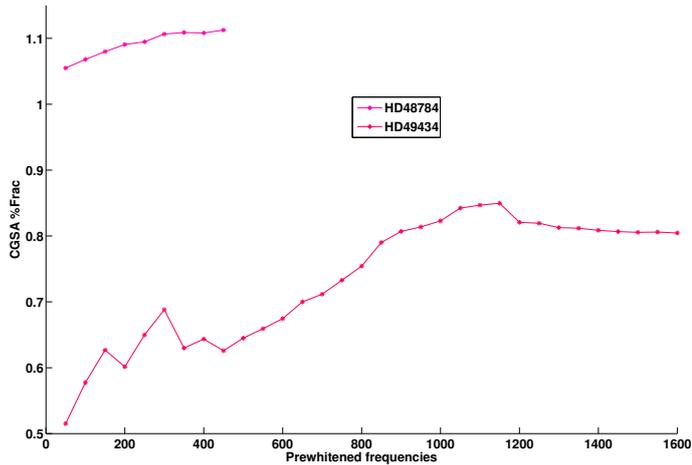}
	\caption{$\% Frac$ CGSA vs pre-whitening step for the Gamma Doradus stars samples. Points are separated by 50 pre-whitening steps.}
	\label{CGSA_vs_step_green_All}
\end{figure}

\begin{figure}
	\centering
		\includegraphics[width=9cm]{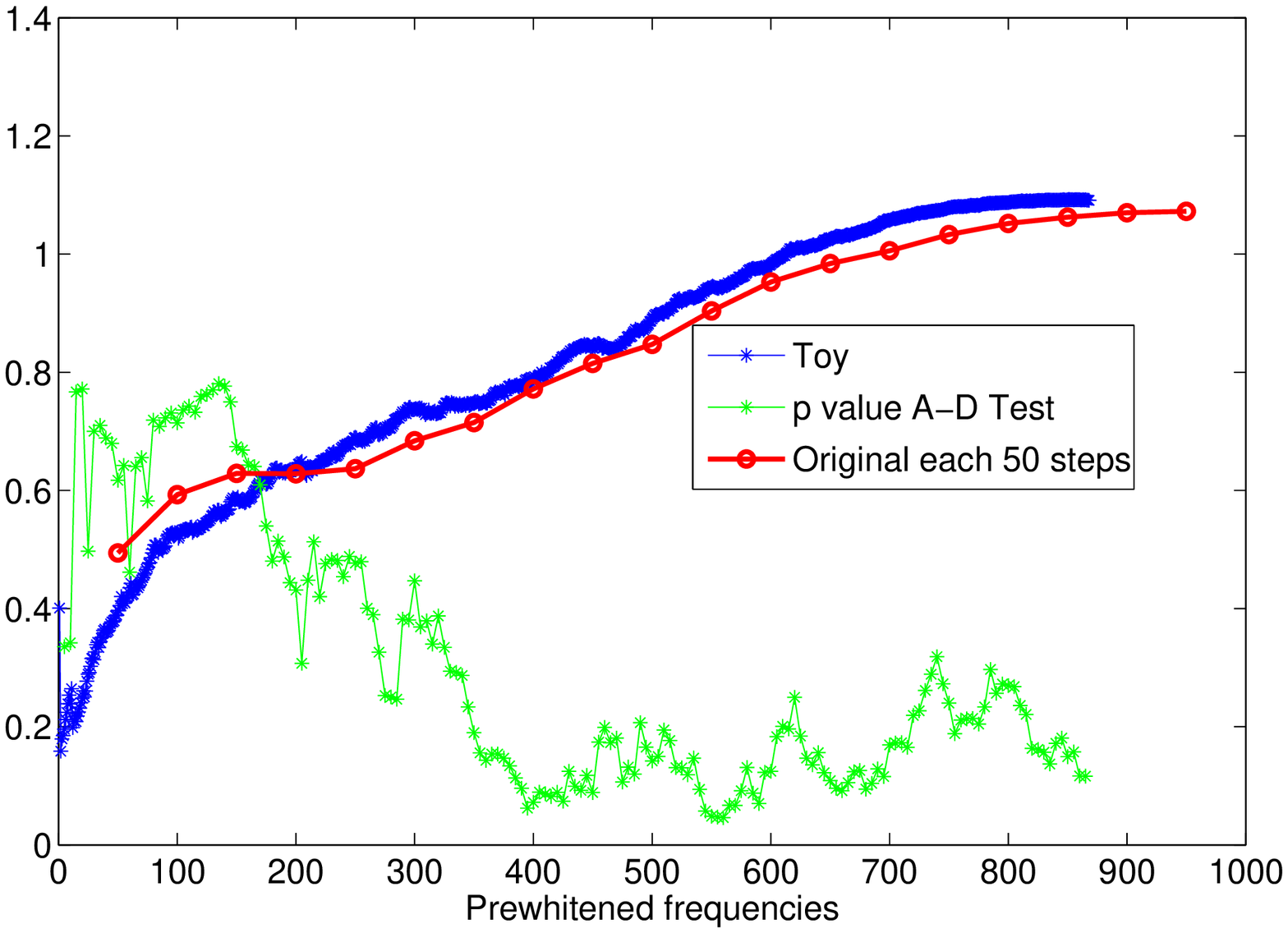}
	\caption{$\% Frac$ CGSA vs pre-whitening step for HD174532 and its toy model.}
	\label{CGSA_vs_step_Toy}
\end{figure}

\section{Conclusions and future prospects}\label{sec:conclusions}
In this paper we tackled one of the most important problems in asteroseismology: a robust and reliable determination of the stellar frequency content. To do so, we have considered two recent results: 1) we have shown in Pascual-Granado et al. 2018 that the pre-whitening method to get stellar pulsation frequencies (which is widely used in the field) is not efficient in most cases, in particular when light curves present effective duty-cycles under 95\%; 2) we have demonstrated that the ultra-high precision photometric light curves of $\delta$ Scuti stars observed with satellites show a fractal behavior (de Franciscis et al. 2018). This unexpected result paved the way to disentangle different phenomena in those stars, e.g. those coming from activity, convection, and the harmonic oscillation modes. \\
In this work we performed a systematic fractal analysis to every single step of the pre-whitening procedure applied to the well-characterized sample of $\delta$ Scuti stars of CoRoT seismo-field. Our basic hypothesis is that light curves are composed by a superposition of proper (harmonic) modes, together with a typical fractal background noise. We applied the CGSA algorithm to discriminate the stochastic fractal power spectra from the harmonics.\\
This work gives support to Pascual-Granado et al. 2015 hypothesis about the fractal origin of the non-analyticity found in the functions underlying the asteroseismic time series observed with CoRoT and Kepler. The present work gives us arguments and tools to tackle this problem. 

\section*{Acknowledgements}

SdF, JPG, JRR, MLM and RG acknowledge funding support from Spanish public funds for research under projects ESP2015-65712-C5-3-R. JCS \& AGH acknowledge funding support from Spanish public funds for research under projects ESP2017-87676-2-2 and ESP2015-65712-C5-5-R. SdF, JPG and RG acknowledge support from the 'Junta de Andaluc\'ia' regional government under project 2012-P12-TIC-2469. JCS also acknowledges funding support from project RYC-2012-09913 under the 'Ram\'on y Cajal' program of the Spanish MINECO. Based on data from the COROT Archive at CAB.


\clearpage
\appendix

\bibliographystyle{mnras}
\bibliography{lamiabib} 
\end{document}